%% file: main.tex
\documentclass[a4paper, 12pt,oneside]{article}
\usepackage[top=3cm, bottom=3cm, left=2.5cm, right=2.5cm]{geometry}
\usepackage{graphicx}
\usepackage[skip=10pt plus1pt, indent=20pt]{parskip}
\usepackage{amsmath}
\usepackage{amsthm}
\usepackage{bbold}
\usepackage{amssymb}
\usepackage{authblk}
\usepackage{amsfonts}
\usepackage{mathrsfs}
\usepackage{bigints}
\usepackage{braket}
\usepackage{siunitx}
\usepackage{makecell}
\usepackage{physics}
\usepackage{enumitem}
\usepackage{float}
\usepackage{blkarray}
\usepackage{braket}
\usepackage{booktabs}
\usepackage{pgfplots}
\usepackage{pgfplotstable}
\pgfplotsset{compat=1.15} 
\usepackage{multirow}
\usepackage{hyperref}
\usepackage{multicol}
\usepackage{booktabs}
\usepackage{tcolorbox}
\usepackage{lipsum}
\usepackage{xcolor}
\usepackage{wrapfig}
\usepackage{subcaption}
\usepackage{cite}
\usepackage{longtable}
\usepackage{appendix}
\usepackage{mathtools}
\usepackage{csvsimple}
\usepackage{relsize}
\usepackage{graphicx} % Required for inserting images
\usepackage{tikz}
\usetikzlibrary{quantikz2}
\usetikzlibrary{matrix}

\input{commands}

\title{Specifying the operational meaning \\ of quantum reference frames}
\author[1]{Augustin Vanrietvelde}
\affil[1]{Télécom Paris, Institut Polytechnique de Paris, Inria Saclay, Palaiseau, France}
\date{}

\begin{document}

\maketitle

% Marseille, 10 juin 2026

\begin{abstract}
    In their strongest usage, quantum reference frames have been described as referring to `the measurements performed by a superposed lab', `the perspective of a quantum particle', `the point of view of a superposed observer', etc. While exciting, these operational proposals have remained brief and ambiguous, leading to misinterpretations and criticism. Here, we provide a detailed specification and defense of the notion of a position-superposed lab or observer. We argue that this requires no exotic claims about quantum physics and raises no greater interpretive difficulty than ordinary quantum measurements. We then derive several consequences of taking this operational meaning seriously. We stress that the position-superposed observers that define quantum references frames are different from, and considerably less proble\-matic than, the outcome-superposed observers considered in Wigners' friend scenarios. In particular, we show that outcomes obtained by a position-superposed observer may (without decohering the superposition) be broadcast to a well-localised one, in contrast with Wigner's friend scenarios, which require the outcomes to remain internal to the system at hand. Finally, we defend the possibility to `roleplay' a quantum reference frame from a classical reference frame.
\end{abstract}

\section{Introduction}

% Interest in QRFs

The idea of quantum reference frames\footnote{The word `quantum reference frames' has been used with various meanings. Here, we are concerned with the strongest of those meanings, in which a QRF is explicitly linked to the perspective of potentially superposed observers, as described in the present paper. This is sometimes called an `internal' or `perspectival' QRF. The defining reference is Ref.~\cite{Giacomini:2017zju} (with Ref.~\cite{Angelo_2011} an important precursor), with further elaboration and extension in e.g.\  Refs.\ \cite{Giacomini:2018gxh, Ballesteros:2020lgl, Giacomini:2021gei, Castro-Ruiz:2021vnq}. Besides this approach, one can broadly identify several classes of proposed meanings for QRFs: as symmetry-breaking resources \cite{Smith:2018gas, Spekkens:2008acy, Gour:2009hng, Bartlett:2006tzx}, as ways to relativize observables \cite{Loveridge:2016tnh, Loveridge:2017pcv, Carette:2023wpz, Glowacki:2023nnf}, or as choices of relational perspectives within gauge systems \cite{Vanrietvelde:2018pgb, delaHamette:2020dyi, delaHamette:2021oex, Hoehn:2019fsy,Hoehn:2021flk,  Krumm:2020fws, Hoehn:2023ehz} (the latter being closer to the idea of internal QRFs). While distinct, these meanings display complex interdependencies with one another. For comparisons of different approaches, see Refs.~\cite{Krumm:2020fws, Castro-Ruiz:2025yvi, Doat:2025klp}.} (QRFs) is to ask:
\begin{center}
    \textit{`How would the world look like \\ from the perspective of a superposed observer?'}
\end{center}
This question is exciting because it holds the promise of re-enacting the spectacular \textit{tour de force} of special and general relativity: 1/ introducing seemingly eccentric perspectives (that of an observer in motion with respect to the aether, that of a deformed coordinate system); 2/ positing that these are in fact just as good as our own, and that our initial dismissal of them was rooted in baseless prejudice; 3/ building on this radical principle a successful physical theory. For QRFs, the hope is to eventually serve as a conceptual cornerstone of quantum gravity.

% can be defined as `references frames corresponding to a quantum system', as `how a quantum observer describes the world', as `the rest frame of a superposed particle', and so on. e

% Problem of their operational meaning / what I mean by operational meaning
Yet, promising as it seems, this idea comes with a major conceptual liability. In the build-up to special relativity, it is straightforward to picture an observer performing measurements from within a moving rocket. But what does one mean exactly by a `superposed observer' or -- another typical phrase -- `the frame/perspective of a quantum particle'? After all, a particle cannot `see' or measure anything. As for the notion of a `superposed observer', it might trigger a lingering discomfort: isn't it a basic tenet of quantum theory -- at least in some interpretations, typically Bohr's -- that an observer is \textit{by definition} classical? Haven't `superposed observers' been shown time and again to bring about a whole zoo of troublesome Wigner's-friend-like paradoxes, well before gravity even comes into play?

The pivotal issue here is that of the \textit{operational meaning} of QRFs. Various QRF frameworks give us mathematical ways to translate between the outcome probabilities of e.g.\ `measurements made from the perspective of particle $A$' and `measurements made from the perspective of particle $B$'. But they do not make it clear how such measurements are to be performed in practice, leaving themselves vulnerable to the accusation of being mere mathematical games, with no bearing on `physics in the lab', in a stark contrast with the transparent meaning of e.g.\ relativistic reference frames.

Previous literature -- for instance, Refs.~\cite{Giacomini:2017zju, Giacomini:2018gxh} -- has sometimes alluded to an operational meaning, which we may summarise (in the simplest case of QRFs for spatial position) as the following:
\begin{center}
    \textit{Physical quantities described from within a quantum reference frame (of spatial position) are the ones that would be measured by measurements performed within a corresponding lab whose absolute position may be superposed.}
\end{center}
These have however remained passing allusions, as opposed to a clear specification of the idea of superposed labs. As such, they have not prevented the idea of QRFs from garnering criticism, exemplified by Ref.\ \cite{Adlam:2022zar}.\footnote{In addition to this reference, much of the criticism we have in mind has been expressed orally. Note that Ref.\ \cite{Adlam:2022zar} also expresses criticism of the Page-Wootters approach to define temporal QRFs, which it is not our aim to discuss here.}

% Proposal here

The goal of this paper is to resolve this situation, by providing a thorough explanation, defense, and discussion of the above proposal for an operational meaning of QRFs as referring to measurements performed by superposed observers or labs. Our first contribution is to specify this proposal into an unambiguous and detailed one.

Our second contribution, intertwined with the first, is a detailed defense of such a notion. We argue that, although obviously outlandish in current experimental practice, \textit{these observers are physically credible}: they rely neither on a particular interpretation of quantum theory, nor on dubious stretches of it. And more specifically, we take great care to stress that \textit{these `superposed observers' are not the ones that appear in Wigner's friend paradoxes}. Wigner's-friend-like observers are described (by an external observer) as holding a superposition of measurement outcomes; the observers defining QRFs, while being in a superposition of absolute spatial position (or of another specific quantity), stand outside of the Heisenberg cut as far as their measurement registers' internal states are concerned, and are described (even by an external observer) as holding definite measurement outcomes. Said in another way, making operational sense of QRFs requires no particular stance on the measurement problem. 

% Consequences of this proposal

Our third contribution is the discussion of a few important consequences of taking these observers seriously. First, we stress that a position-superposed laboratory necessarily stores its measurement outcomes in position-superposed registers, but that this is not physically problematic. Second, we prove (by introducing a concrete communication scheme) that measurement outcomes collected within a position-superposed QRF may be communicated to an agent sitting in a well-localised reference frame (without collapsing the original QRF's superposition of spatial positions), and even sent away to infinity -- which also proves that measurements from within a QRF cannot be subject to Wigner's friend paradoxes. Third, leveraging an analogy with the fact that relativistic time-dilation was proven without the use of labs going at relativistic speeds, we argue that superposed labs may, under suitable physical assumptions, be `roleplayed' from within a standard reference frame, opening the door to more realistic checks of QRF frameworks.

% Not particularly new?

% (It should be stressed that the operational meaning discussed here is not wholly novel; my feeling is that it has been implicit -- or sometimes ambiguously alluded to -- through much of the perspectival strand of QRF literature. I do not expect it to come as a surprise to most people working on that strand. Still, I expect that there is added value in specifying it clearly and in detail, arguing in favour of its credibility, and discussing its consequences. I have often encountered criticisms of QRFs resting on an incorrect understanding of this crucial ingredient.)

% The tone here?

In this paper we will focus on the simplest example of QRFs: those that are superposed in spatial position. These are quantum analogues to the simplest kind of classical reference frames, those for position, determining an origin of coordinates in (non-relativistic) space. Many more elaborate types of QRFs have been proposed: for rotations, time, relativistic coordinates, coordinate systems, and so on. While the extension of our proposal to these might involve subtleties (especially in the case of temporal reference frames), we expect it to work on a similar basis.

% A note about the style of this paper. Remarkably, it does not feature any equation (at least in the main text). There are two related reasons for this. First, the point of this paper is to discuss concepts; equations would bring little additional clarity and draw the attention away from our real proposed contribution. Proposals for detailed computational frameworks describing QRFs can be found in the rich existing literature. Second, these frameworks have been shown to be mathematically inequivalent, and we would like our proposed operational interpretation to remain agnostic with respect to these various mathematical formalisms.

\section{The proposal} \label{sec: proposal}
\subsection{Arbitrarily complex systems in a superposition of positions}
Take a particle – say, an atom; it can be in a superposition of two (absolute\footnote{\label{foot:relativity}Of course, the whole point of the QRF program is to eventually deny physical meaning to this `absolute position' of the particle, and to posit that it being in a superposition is a merely perspectival statement. But to get there, we need to start from a situation in which this absolute position is taken seriously. This is similar to how any symmetry has to be described on the backdrop of a (fictional) situation unconstrained by the symmetry.}) positions,
\be \ket{\psi} = \frac{1}{\sqrt{2}} \Big( \ket{x_0} + \ket{x_1} \Big) \, . \ee 
Now, make that two atoms, bound into a molecule. This molecule can itself also be in a superposition of two positions; by which we mean that, while the relative positions of each atom with respect to the other are determined by the molecule’s physics – how tight the molecular bound is, whether they are in an excited state, etc –, the molecule's absolute position\footnote{Trough this paper, by a system's `absolute position', we will primarily mean the absolute position of its centre of mass. However, the absolute position of any constituent of the system can play the same role, if this constituent is at a definite position relative to the centre of mass. Therefore, such a position could equally well be taken as `the system's absolute position' without affecting our discussion. We expand on this in Appendix \ref{app: absolute position}.} would be in a superposition. For instance, if the latter superposition is even, the state would be written as
\be \label{eq: even superposition} \ket{\psi} = \frac{1}{\sqrt{2}} \Big( \ket{x_0} + \ket{x_1} \Big)_\ab \otimes \ket{\phi}_\inte \, , \ee 
where we decomposed our degrees of freedom between absolute position and the relative position of one atom with respect to the other. (We call it `int' as an internal degree of freedom of the molecule.) The latter is in an arbitrary state $\ket{\phi}_\inte$, which can in many situations evolve independently of the absolute position. For instance, the molecule could spin around its centre of mass.

Now, nothing forbids us from adding to this molecule a third atom, and a fourth, and so on. We get to an increasingly intricate molecule, with more and more internal degrees of freedom (deformations, local or global rotations, excitations of some of its bounds, and so on).\footnote{In fact, this was already the case of our initial atom: we could have decided to include in our description its internal degrees of freedom, such as the potentially excited state of its electrons.} Still, it is perfectly conceivable that, irrespective of and uncorrelated with the state of these inner degrees of freedom, the molecule’s absolute position is in a superposition. Furthermore, it could remain so – with the `position of the centre of mass’ degree of freedom remaining independent of the internal degrees of freedom – as long as the molecule doesn’t interact with the outside world, or interacts with it in a suitably controlled way. For an even superposition, the state is still of the form (\ref{eq: even superposition}), only with a larger `int' Hilbert space of internal degrees of freedom.

Let’s keep going. We can bring in another molecule, here again in such a way that the composite system’s absolute position is in a superposition. And a third, and a fourth, and so on. We get to a gas of molecules. Its space of internal degrees of freedom gets larger and larger, undergoing more and more elaborate dynamics: collisions, chemical reactions, exchange of light between molecules; all the while its absolute position remains in a superposed state. Of course, at this point, the situation is getting less and less realistic: surely at some point one of these millions of atoms will emit a particle, holding information about its absolute position, which will get lost at infinity, decohering the superposition; or an external photon will hit the system and couple its absolute position with its internal degrees of freedom. Still, as far as fundamental physics is concerned, these are merely contingent events, that a suitable, ridiculously thorough amount of shielding could in principle prevent.\footnote{An interesting question is whether there is a fundamental limit on the size of macroscopic superpositions, typically due to decoherence by an unscreenable gravitational interaction. In the absence of a clear consensus on this question, we will work from the assumption that there is no such limit.}

(However, many internal degrees of freedom of this molecule gas may get decohered, through interaction with external systems. Such interactions have to be of a specific kind, in order not to decohere the system's absolute position, but they exist; we discuss them in Section \ref{sec: broadcasting outcomes}. It is thus possible for most internal degrees of freedom of our molecule gas to be (indistinguishable from) classical ones, just like what would happen in a typical macroscopic system. At this point, we should switch to a density matrix description, which will be of the form
\be \label{eq: even superposition density} \rho = \frac{1}{2} \Big( \ket{x_0} + \ket{x_1} \Big) \Big( \bra{x_0} + \bra{x_1} \Big)_\ab \otimes \sigma_\inte \, , \ee 
where $\sigma$ may be partially decohered.)

You see where this is going. Ultimately, there’s no clear point at which physics tells us to stop. We can add to our system whole solid objects, fluids, biological structures; we can build it up to a grain of sand, a boulder, a racoon, a microscope, a human being, a particle accelerator, a rocket; we can form a planet, equipped with a functional human society, including broadcasting devices, universities, competing laboratories, scientific journals, academic controversies; myriads of internal degrees of freedom -- most of them fully decohered and classical -- doing their thing, all the while this planet’s absolute position remains in a neat superposition. The only restriction is that the planet remains shielded within a perfectly opaque membrane – maybe with a few pinholes, used for carefully controlled interactions with the outside.

\subsection{Observers in a superposition of positions, and their reference frame}
Now, we all know that the notion of a measurement in quantum theory is an ambiguous one, to say the least. Yet, no matter our stance on the measurement problem, we should agree that a system whose degrees of freedom are those of a planet and include thousands of full labs and a whole arxiv repository could reasonably be treated, one way or another, as indeed performing measurements and collecting outcomes. (What’s more, as we shall see later in Section \ref{sec: broadcasting outcomes}, this planet’s measurement outcomes can even be broadcast to the outside world, making them immune to any kind of Wigner’s friend paradox.) There’s nothing stranger in that than in the fact that we, the actual human society on the actual unsuperposed Earth, who are nothing but a myriad of internal degrees of freedom built in the shape of a planet, can be treated as collecting measurement outcomes, too; apart from our interactions with the rest of the galaxy, which arguably are irrelevant to the problem of whether we measure our own local molecules within our own local labs, nothing distinguishes us from the people of a superposed planet.

Note that there can be no way to certify for sure that measurements are indeed being performed in our superposed planet, simply because the measurement problem and the shiftability of the Heisenberg cut make it so that it is \textit{never} possible to say for certain, from the sole formalism of quantum theory, whether a certain measurement has been performed or not. The same ambiguity would apply as well to a description of the actual, non-spatially-superposed planet Earth. In other words, the specification of measurements from a superposed position suffers from the measurement problem to the exact same extent as the specification of \textit{any} measurement, no more, no less.

Importantly, our planet doesn’t have to remain shielded from the outside. At any point, inhabitants could, say, uncover a porthole and let some sunlight in. Of course, if this is done in an uncontrolled way, then this will immediately decohere the planet’s superposition, which would make the situation uninteresting. But the planet's inhabitants can also decide to interact with accomplices in the outside world in a controlled way, so as to collect and broadcast information without affecting its absolute position (in particular without broadcasting it, which would lead to decoherence). For instance, it can shoot carefully aimed particles at a nearby system (another shielded planet, for instance) and collect them for analysis after they bounced back: if this is done in a way that only results in a momentum transfer between the two planets, it will not have broadcast their positions. Thus, as long as it respects some rules not to destroy its superposition, this planet can look at the outside world.

This finally leads us to the idea of quantum reference frames. Since Galileo (or even before) we’ve known that our position affects how we see the world: our measurements are performed relative to a  \textit{reference frame}, a position taken to be the origin (typically, ours, or that of our measurement device). The same world will be described differently from different reference frames. But we’ve now unlocked a new reference frame: that of an observer (an inhabitant of the planet) who – with respect to a prior reference frame – is not in any given position, but in a superposition of two (or even more). As we’ve hopefully managed to argue, this is as good an observer as any other, collecting outcomes just like we do. And we end up with the central and fascinating question of quantum reference frames:

\begin{center}
    \textit{How would such a position-superposed observer see the world?}
\end{center}
together with many more that go with it: is there a map transforming her perspective into ours?  Is physics the same from her perspective? Is there something to be learned from the fact that physics is indeed invariant when we go to such a perspective – in analogy with how there was so much to be learned from the invariance of physics under Lorentz transformations? How does she see entanglement and superpositions herself? Shouldn’t it be said that, from her perspective, it is we who are superposed?

What’s more, our example was based on superpositions of position, but a similar story could be told in terms of other physical quantities: we can also conceive of observers whose velocity, angular orientation or momentum, or clocks’ ticks are superposed with respect to our own point of view, and ask how they would perceive physics. The story above can broadly be told similarly in those cases (even though there would be subtleties to keep in mind, especially for the case of clocks).

\subsection{The operational meaning of quantum reference frames}
We can now get back to the operational meaning of quantum reference frames, namely, the question of how quantities that our theory deems to be `described from within a quantum reference frame’ can be measured through lab operations. As discussed in the introduction, we simply take it to be the one already alluded to at various points in previous literature, namely:
\begin{center}
    \textit{Physical quantities described from within a quantum reference frame (of spatial position) are the ones that would be measured by measurements performed within a corresponding lab whose absolute position may be superposed.}
\end{center}
Analogous proposals can of course be made about quantum reference frames with respect not to position, but to another quantity, such as velocity, angular momentum, time, and so on.

As discussed in the introduction, the contribution we aimed to have here is, through the previous sections, to have made this meaning unambiguous and credible, by carefully introducing labs in a superposition of position and how they naturally perform measurements.

This proposal has the advantage of dispelling the worries created by other phrases commonly used, such as `the quantum reference frame of a particle’, which will typically trigger pushback along the lines of `but a particle does not measure anything’. In the light of the above proposal, we would say that `the quantum reference frame of a particle’ is an illustrative shortcut to refer to `the reference frame of a lab whose (potentially superposed) position would be that of the particle’. This is nothing really new: in a special relativity course, it is commonplace to refer to `the reference frame of a particle’ (for instance, a muon going close the speed of light). What one means by that phrase is that, even though the muon has no clocks and rods available within its internal degrees of freedom, it is enlightening to consider a hypothetical rocket going at the same speed, and how observers within such a rocket would experience physics.

\section{Important features of the proposal} \label{sec: features}
Let us discuss critical consequences of the proposed operational meaning of QRFs. These are counter-intuitive, but necessarily arise if one decides to uphold it. Furthermore, we will comment on how their counter-intuitiveness is merely superficial, and how they are much more palatable than would seem at first sight.
\subsection{The position of the registers is superposed, too}
A quantum measurement can always be described as the update of one or several physical registers. These registers will typically be the pointers of measurement devices, the pages of lab notebooks, and ultimately screens displaying the corresponding arxiv preprint. In our example of a measurement performed on a planet in a superposition of positions, any of these registers, located on the planet, is also in a superposition of (absolute) positions, typically entangled with the absolute position of the centre of mass.

Indeed, consider for simplicity the case of a single register, at a well-defined relative position $y$ from the centre of mass. Let us then move, following Appendix \ref{app: absolute position}, to a description of our system divided between the register's absolute position (denoted `RAP') and internal degrees of freedom. If the planet has remained in a pure state, for an even superposition this state can be written as
\be \label{eq: even superposition with reg} \ket{\psi} = \frac{1}{\sqrt{2}} \Big(\ket{x_0+y} + \ket{x_1+y}\Big)_\RAP \otimes \ket{{\phi}}_\inte \, , \ee 
For a planet with some internal degrees of freedom decohered we simply get the more cumbersome doubled form
\be \label{eq: even superposition with reg dens} \begin{split}
    \ket{\psi} = \frac{1}{2} \Big(\ket{x_0+y} + \ket{x_1+y}\Big) \Big(\bra{x_0+y} + \bra{x_1+y} \Big)_\RAP \otimes {\sigma}_\inte \, . 
\end{split} \ee 

This might, at first sight, spook us: from our external, `well-localised’ perspective, we do not see such a register as a well-localised degree of freedom containing the measurement outcome. Instead, we see a superposition of `the register at $x_0+y$, containing the measurement outcome' and `the register at $x_1+y$, containing the measurement outcome'. This might lead us to the conclusion that no measurement has happened, and that everything is still in a superposition.\footnote{Worries of this kind are for instance expressed in Ref.\ \cite{Adlam:2022zar}'s Section 5.}

Such a worry is unwarranted, however. Indeed, what matters is not the register’s spatial position, but its internal state (which is part of $\sigma_\inte$ in (\ref{eq: even superposition with reg dens})); it is in this internal state that the measurement outcome has been recorded. And it is perfectly feasible, should we decide to place ourselves out of the Heisenberg cut, to indeed hold that the measurement has been performed, i.e., to apply the Born rule and collapse the register’s internal state to a specific classical outcome. This only amounts to updating $\sigma_\inte$, not the left-hand factor of (\ref{eq: even superposition with reg dens}). The register's internal state and its absolute spatial position are different degrees of freedom, and the former undergoing a collapse is perfectly compatible with the latter remaining superposed.

Thus, if we consider position-superposed observers, we necessarily have to bite the bullet of position-superposed outcome registers. But it is a bitable bullet: what matters in an outcome register is not its position, but its internal state encoding the outcome obtained.

\subsection{Measurement outcomes can be broadcast to other reference frames}
\label{sec: broadcasting outcomes}
A critical feature of measurements is that their outcomes should be copyable and broadcastable to other registers at will. It should even be possible for us to copy them onto a register then lose this register (for instance if this register is the internal state of a photon that escapes to infinity). Some decoherence-based interpretations of quantum theory, à la Zurek, would even hold that this copying-then-losing-the-copy is what \textit{defines} a measurement process. On the other hand, a measurement process whose outcome is never broadcast to the external world and remains confined solely to internal registers is to be regarded as, at the least, highly suspicious. For instance, it will typically lay itself open to the whole array of Wigner’s friend-type paradoxes.

It would thus be highly worrying if measurement outcomes obtained within a quantum reference frame could not be broadcast. Fortunately, this is \textit{not} the case: \textit{measurement outcomes obtained from within a quantum reference frame may be communicated out, to registers (or agents) sitting in a `well-localised’ reference frame} – without decohering the original system’s superposition of positions. A superposed rocket could very well communicate its outcomes to us, `well-localised’ people, or for instance send copies of those out to infinity, while remaining in a superposition.

Let us see how. Suppose we have two agents: Alice, sitting in a rocket, and Eve, standing outside. The centre of mass of Alice's rocket lies (with respect to Eve’s reference frame) at a superposition of two positions $x_0$ and $x_1$. Alice obtained a measurement outcome $s$ that she wants to communicate to Eve. We suppose that Eve knows that Alice's absolute position is an even superposition of $x_0$ and $x_1$.\footnote{In fact, our protocol works even if Eve only knows that Alice is lying at any, arbitrarily weighted, superposition of $x_0$ and $x_1$, including the possibility that she is well-located at one of these two positions.}

To simplify, we will assume that Alice's register lies at the rocket's centre of mass. As we discussed, this measurement outcome is initially written in a physical register located within Alice’s rocket, with the overall state being
\be \label{eq: before broadcasting} \begin{split}
    \rho_{\textrm{initial}} = \frac{1}{2} \Big(\ket{x_0} + \ket{x_1}\Big) \Big(\bra{x_0} + \bra{x_1} \Big)_\RAP \otimes \ket{s}\bra{s}_{\reg_A} \otimes {\sigma}_\inte \, . 
\end{split} \ee 
(`$\reg_A$' stands for the internal state of Alice's register.) A naively designed interaction between Alice’s and Eve’s registers runs a risk of accidentally disclosing Alice’s absolute position and collapsing her superposition.

A strategy to perform the broadcasting without disclosing Alice’s position is the following. Eve takes her register and first places it next to $x_0$. At this point, she triggers an interaction (for instance by turning on a potential) through which, if Alice’s register is located in the vicinity (i.e.\ if she is located at $x_0$), then its state gets copied to Eve’s register. If Alice’s register is not in the vicinity (i.e.\ if she is located at $x_1$), then the two registers are too far away and no interaction takes place. After this interaction, the state has become
\be \label{eq: intermediate broadcasting} \begin{split}
    \rho_{\interm} = \ket{\Psi_\interm}\bra{\Psi_\interm} \otimes \ket{s}\bra{s}_{\reg_A} \otimes {\sigma}_\inte \, . 
\end{split} \ee 
where
\be \label{eq: intermediate broadcasting Psi} \ket{\Psi_\interm} = \frac{1}{\sqrt{2}} \Big(\ket{x_0}_\RAP \ket{s}_{\reg_E} + \ket{x_1}_\RAP \ket{0}_{\reg_E}\Big) \ee 
Note how, after this interaction, the internal state of Eve’s register may get temporarily entangled with Alice’s position.

Eve then moves her register over to $x_1$ – taking care not to measure its internal state yet – then triggers the same interaction a second time. Once this second interaction has taken place, the state becomes 
\be \label{eq: final broadcasting} \begin{split}
    \rho_{\fin} = \ket{\Psi_\fin}\bra{\Psi_\fin} \otimes \ket{s}\bra{s}_{\reg_A} \otimes {\sigma}_\inte \, . 
\end{split} \ee 
where
\be \label{eq: final broadcasting Psi} \ket{\Psi_\fin} = \frac{1}{2} \Big(\ket{x_0}  + \ket{x_1} \Big)_\RAP \ket{s}_{\reg_E} \, .\ee 
Eve's register is in a product state with Alice’s position, and now contains Alice’s measurement outcome. Alice’s measurement outcome has thus been broadcast to Eve, who can furthermore broadcast it further to all her well-localised friends (or to position-superposed ones, through a symmetric procedure), send a copy away to infinity, and so on.

On a final note, the interaction could even be designed in such a way that Alice's register gets erased while its internal state is copied to Eve's. In such a situation, Alice herself doesn’t even keep a copy of her outcome.

\subsection{Superposed labs may be roleplayed}
Our proposed operational way of making sense of quantum reference frames, with its reference to superposed labs, is obviously unfeasible in practice: maintaining even a mesoscopic system in a superposition of positions is atrociously difficult. However, we want to argue (in line with previous arguments made in Ref.~\cite{Doat:2025klp}) that this does not make it impossible for us to collect experimental data pertaining to a quantum reference frame. We simply have to accept to roleplay a bit, something we routinely do in other contexts.

The relevant parallel is, once again, with special relativity. In the latter, an inertial frame is often – rightfully, in my opinion – operationally interpreted as referring to the measurements obtained by agents employing rods and clocks from within a system going at a certain velocity. Yet, demonstrations of relativistic effects, such as time dilation, by use of \textit{bona fide} clocks are a relatively recent performance, unlocked in the 70s by the development of high-precision atomic clocks; and the predictions of special relativity were held to have been experimentally confirmed long before these demonstrations took place. This is because earlier experiments had already roleplayed `agents going at relativistic velocities’.

The Ives-Stilwell experiment, usually taken to be the first demonstration of time dilation, is for instance based on the measurement of a relativistic Doppler effect in the frequency of the light emitted by atoms in a beam. The connection to time dilation is that this Doppler effect may be ascribed to the atoms’ transition frequency running slow – amounting to a dilation of the atom’s `internal clock’. But of course, calling transition frequencies an `internal clock’ is a stretch, to say the least. The – often implicit – way to make the connection is to say that, \textit{if there were an observer (endowed with a pretty good set of lab equipment) at rest with respect to the atom, she could use this transition frequency as a frequency reference and ultimately as a clock}. We see that the observer here is a mere potentiality. But since our theory of physics tells us that nothing prevents this potentiality, we are entitled to reason in that way. As already discussed in Ref.\ \cite{Doat:2025klp}, the same can be said about the second historical proof of time dilation, the sea-level detection of muons, whose status as a proof of time dilation is based on the idea that the atmospheric disintegration of a cosmic muon would amount to a basic clock for a – non-existent – observer at rest with respect to it.

In interpretations of those two experiments, the critical and often unspoken assumption is the idea that one can refer to an agent that would operate within a system’s reference frame – that would be at rest with respect to the particle at hand – even if there is in fact no such agent. The fact that the agent could exist – that nothing in the laws of physics forbids it in principle – is sufficient.

The same can be said about quantum reference frames. Suppose we have a small system – even a single particle – in a superposition, and sound physical reasons to uphold that a certain quantity $Q$ that we (in our boring, well-localised lab) can measure on it would always have the same value as the quantity $Q’$ that a (non-existent) agent well-localised with respect to the particle (and therefore delocalised with respect to us) would measure. Then we are entitled to claim that measurements of $Q$ are the same as measurements of $Q’$. The important consequence is that, under specific circumstances to be determined by our knowledge of physics, \textit{we are allowed to test predictions made by quantum reference frames theories by performing certain measurements in our own classical reference frame} – in the same way that relativistic time dilation was first confirmed by frequency measurements ultimately performed in a boring, non-relativistic frame.\footnote{An interesting concrete instance of a measurement ultimately performed in a classical lab but arguably pertaining to a QRF is that of the `quantum ruler' introduced in Ref.~\cite{Wang:2023koz}.}

In conclusion, we don’t have to wait for a position-superposed human being to exist in order to probe predictions about quantum reference frames. This would be the same as waiting for a human being to move at 99\% the speed of light to check time-dilation.

% \section{Replies to a few potential criticisms (or: Frequently Asked Questions)}
% Obviously, this proposal is bound to trigger many criticisms; here I will pre-emptively reply to a few, drawing mostly from what I was faced with in personal conversations. This will take the form of answers to typical questions.\footnote{We are getting closer to a Galileo-style `philosophical dialogue’ between a character called Smartus and another called Stupidus. I will certainly not cross that line, out of respect for my esteemed colleagues.}
% \subsection{These are not the same superposed observers as in Wigner’s friend paradoxes}
% Q: 

\section{Conclusion}

This paper made three main contributions: an unambiguous specification of the operational meaning of QRFs as alluded to in previous literature; a defense of its physical relevance; and a discussion of several important consequences that stem from taking this operational meaning seriously.

To avoid misunderstandings, it may be important to mention a few things that this paper neither did nor aimed to do. First, the main goal of the QRF program is to eventually argue that `a quantum reference frame is as good a reference frame as any other' -- for instance through the introduction of some kind of extended covariance principle, positing the invariance of physical laws under QRF transformations. We did not engage with this task: our work should be seen as a prerequisite to making it meaningful, by specifying the operational meaning of QRFs in the first place. This is why we allowed ourselves to talk about a lab `being in a superposition', even though the eventual aim of the research program is to argue that this `being in a superposition' is only a perspectival statement (see also footnote \ref{foot:relativity}).

Second, our goal was not to give a full mathematical account of QRFs or transformations between those, as can be seen in the fact that the maths in this paper remained at a homeopathic level, and serving mostly as illustration. Our point here was first and foremost conceptual, serving as a complement to the rich and very detailed mathematical formalisms of which the QRF literature has no shortage. Finally, we did not make any claims as to the answer to our inaugural question, that of `how the world would look like from the perspective of a superposed observer'. Our goal was merely to argue that this question makes sense, because a superposed observer can be conceived of.

On the question of follow-ups to our work, the most pressing would be an extension of our discussion to \textit{temporal} QRFs, linked to `superposed clocks'. Indeed, these lie at the heart of the current controversy on possible implementations of indefinite causal order \cite{moller2020gravitational, oreshkov2019time, delaHamette:2022cka, Ormrod2023, Vilasini2025, Apadula:2026orf}, and more generally lead to both the most promising and most disputed consequences of the QRF program -- it is telling, for instance, that Ref.~\cite{Adlam:2022zar} pairs its criticism of QRFs with criticism of the Page-Wootters formalism, which anchors some approaches to temporal QRFs.

For QRFs, the most important challenges lie ahead, the main one being to deliver on the implicit promise we mentioned in the introduction: being a catalyst for new physics. While there have recently been interesting incursions of QRFs into quantum gravity \cite{Araujo-Regado:2024dpr, Carrozza:2021gju, Carrozza:2022xut, Kabel:2023jve, Belenchia:2018szb, Kabel:2022cje, Kabel:2024lzr, Cepollaro:2021ccc, Chen:2024xvm}, it remains to be seen whether superposed rockets can play as seminal a role for our future theories of the world as the classical ones did for relativity.

\section*{Acknowledgements}
It is a pleasure to thank Luca Apadula, Guilhem Doat, Thomas Galley, and Nick Ormrod for comments on an early draft of this paper. Special thanks go to Tein van der Lugt, who (through his skepticism) triggered most of the ideas presented here during a walk through Paris in October 2022.
% Many thanks also to Shashaank Khanna and Ravi Kunjwal for inviting me over to Marseille and then not being there, giving me time to write the bulk of this paper from a café in the city centre.

The author is supported by the STeP2 grant (ANR-22-EXES-0013) of Agence Nationale de la Recherche (ANR), the PEPR integrated project EPiQ (ANR-22-PETQ-0007) as part of Plan France 2030, the ANR grant TaQC (ANR-22-CE47-0012), and the ID \#62312 grant from the John Templeton Foundation, as part of the \href{https://www.templeton.org/grant/the-quantum-information-structure-of-spacetime-qiss-second-phase}{‘The Quantum Information Structure of Spacetime’ Project (QISS)}.

\bibliographystyle{utphys}
\bibliography{refs}

\appendix
\section{Trading one absolute position of reference for another} \label{app: absolute position}
In equations such as (\ref{eq: even superposition}), we describe our systems in a way that splits their degrees of freedom between the absolute position of the centre of mass on the one hand, and internal degrees of freedom on the other hand. In this appendix we want to stress that this is to a good extent conventional: the centre of mass is not the only degree of freedom that can play the role of the `absolute position of reference' being superposed.

Indeed, consider any constituent $R$ of the system (for instance, a particle of our choice, or the centre of mass of a subsystem of our choice), and suppose that it lies at a well-defined relative position $y$ to the centre of mass. Then we can decide to adopt a description in which our degrees of freedom are split, not between the absolute position of the centre of mass on the one hand, and internal degrees of freedom on the other hand; but between the absolute position of $R$ on the one hand, and internal degrees of freedom on the other hand. If the state in the original description is\footnote{Here as well as in the next equation, $\ket{\phi}_\inte$ contains, among other things, the information that the relative position between the centre of mass and $R$ is $y$.}
\be \ket{\psi} = \frac{1}{\sqrt{2}} \Big( \ket{x_0} + \ket{x_1} \Big)_\ab \otimes \ket{\phi}_\inte \, , \ee 
then in the new description it is
\be \ket{\psi} = \frac{1}{\sqrt{2}} \Big( \ket{x_0+y} + \ket{x_1+y} \Big)_\ab \otimes \ket{\phi}_\inte \, .  \ee
The same move could be done with descriptions in terms of density matrices. (Of course, since we have traded one way of splitting our degrees of freedom with another, the meanings of the `abs' and `int' parts of the state in these two equations are different.)

The critical point to notice here is that, for any constituent $R$ whose relative position to the centre of mass is well-defined, the statements `the absolute position of the centre of mass is superposed' and `the absolute position of $R$ is superposed' are equivalent ones. This motivates our choice in the main text to simply talk about `a system's absolute position' being superposed: while this should primarily be interpreted as a statement about the absolute position of its centre of mass, it is in fact equivalent to the same statement being made about the absolute positions of many of its constituents.
\end{document}

%% file: commands.tex
\def\be{\begin{equation}}
\def\ee{\end{equation}}
\def\ba{\begin{align}}
\def\ea{\end{align}}

\DeclareTextFontCommand{\texttt}{\ttfamily\upshape}
\DeclareTextFontCommand{\textrm}{\rmfamily\upshape}

\newcommand{\ab}{{\textrm{abs}}}
\newcommand{\inte}{{\textrm{int}}}
\newcommand{\RAP}{{\textrm{RAP}}}
\newcommand{\reg}{{\textrm{reg}}}
\newcommand{\interm}{{\textrm{intermediate}}}
\newcommand{\fin}{{\textrm{final}}}
\usepackage{stackengine}